\documentclass[12pt]{article}
\usepackage{amsmath, amsfonts,euscript,bbm}
\usepackage{epsfig, cite}
\usepackage{color}
\usepackage{ifthen}
\usepackage{graphicx}
\newboolean{Notes}\setboolean{Notes}{true}
\newcommand{\notes}[1]{\ifthenelse{\boolean{Notes}}{\textcolor{black}{#1 
}}{}}

\newcommand{\mb}[1]{\boldsymbol{#1}}

\newcommand{\Expect}[1]{\left\langle #1 \right\rangle}
\newcommand{\NOrder}[1]{\text{:}#1\text{:}}
\DeclareMathOperator{\GammaFunction}{\Gamma}
\newcommand{\GF}[1]{\GammaFunction\!\left(#1\right)}
\newcommand{\TF}[3]{\Theta_{#1}\bigl(#2\bigl|#3\bigr)}

\def\hf{\frac12}
\def\ap{{\alpha^{\prime}}}
\def\Tr{\text{Tr}\;}
\def\V{\mathcal{V}}
\def\A{\mathcal{A}}
\def\M{\mathcal{M}}

\def\Im{\text{Im}\,}

\newcommand{\sect}[1]{\section{#1}\setcounter{equation}{0}}

\newcommand{\skyp}[1]{}
\newcommand{\half}{{\textstyle \frac{1}{2}}}
\newcommand{\Klmt}{\mbox{K\hspace{-7.8pt}KLM\hspace{-10.35pt}MT}\ }
\newcommand{\klmt}{\mbox{K\hspace{-7.6pt}KLM\hspace{-9.35pt}MT}\ }
\def\gsim{\stackrel{>}{{_\sim}}}
\def\lsim{\stackrel{<}{{_\sim}}}

\begin{document}

\bigskip
\hskip 4.5in\vbox{\baselineskip12pt \hbox{CU-TP-1112}  
\hbox{CLNS 04/1880} 
\hbox{NSF-KITP-04-55} }
\bigskip\bigskip

\centerline{\Large Collisions of Cosmic F- and D-strings}
\bigskip
\bigskip
\bigskip
\centerline{{\bf Mark G. Jackson}\footnote{Address after Sept. 1, 2004:
Theoretical Astrophysics,Ê Fermilab, Batavia, IL 60510}}
\medskip
\centerline{Department of Physics}
\centerline{Columbia Univ.}
\centerline{New York, NY 10027}
\centerline{\it markj@phys.columbia.edu}
\bigskip
\centerline{{\bf Nicholas T. Jones}\footnote{Address after Sept. 1,  
2004: University of Amsterdam
}}
\medskip
\centerline{Laboratory for Elementary-Particle Physics}
\centerline{Cornell University}
\centerline{Ithaca, NY 14853}
\centerline{\it nick.jones@cornell.edu}
\bigskip
\centerline{\bf Joseph Polchinski}
\medskip
\centerline{Kavli Institute for Theoretical Physics}
\centerline{University of California}
\centerline{Santa Barbara, CA\ \ 93106-4030, USA}
\centerline{\it joep@kitp.ucsb.edu}
\bigskip
\bigskip

\begin{abstract}
Recent work suggests that fundamental and Dirichlet strings, and their  
$(p,q)$ bound states, may be observed as cosmic strings.  The evolution  
of cosmic string networks, and therefore their observational signals,  
depends on what happens when two strings collide.  We study this in  
string perturbation theory for collisions between all possible pairs of  
strings; different cases involve sphere, disk, and annulus amplitudes.   
The result also depends on the details of compactification; the  
dependence on ratios of scales is only logarithmic, but this is still  
numerically important.  We study a range of models and parameters, and  
find that in most cases these strings can be distinguished from cosmic  
strings that arise as gauge theory solitons.
\end{abstract}

\newpage
\baselineskip=18pt

\sect{Introduction}

The observation of fundamental strings of cosmic size would be a  
spectacular window into short-distance physics.  The existence of such  
cosmic fundamental strings in conventional Planck-scale string models  
is unlikely for several reasons~\cite{witten}: they are unstable either  
to breakage or to confinement by domain walls, and even if stable they  
would be removed by inflation.  However, in lower scale models based on  
large or warped compact dimensions, cosmic fundamental strings may  
indeed exist, as well as cosmic strings arising from D-strings and  
wrapped D-, NS-, and M-branes.  These can be produced after  
inflation~[2-7], and in some models they are stable or at least  
metastable on cosmic time scales~\cite{CMP,LT}.\footnote{There is a  
discrepancy between refs.~\cite{CMP} and~\cite{LT}, which we will not  
address here.}

The observational signatures of cosmic strings depend on the detailed  
evolution of the string network from its creation to  
today~\cite{cosmicstrings}.  This evolution in turn depends on what the  
strings do when they collide.  For two strings of the same type, there  
are two obvious possibilities: they may pass simply pass through one  
another, or they may reconnect (intercommute) as in figure~1.
\begin{figure}[h]
\begin{center}
\leavevmode
\epsfbox{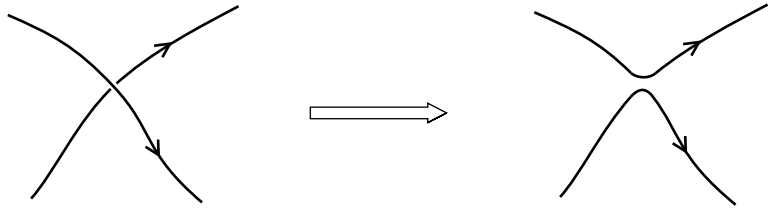}
\end{center}
\caption{When two strings of the same type cross, they can reconnect  
(intercommute).}
\end{figure}
Reconnection contributes to the decay of string networks, allowing  
large loops to break into smaller ones.
For the usual cosmic strings, which arise as solitons (magnetic flux  
tubes) in gauge theories, the reconnection probability is essentially  
one~\cite{matz}.  For fundamental  
strings, however, the reconnection probability is of order $g_{\rm  
s}^2$, and so can be much less than one.  This reduced probability will  
lead to an increased density of strings~\cite{kib}, and so to the  
enhancement of some signatures.

{Optimistically, we may envision an era of precision cosmic string  
cosmology,} when we will distinguish fundamental strings from solitonic  
strings via their intercommutation properties.
In this paper we prepare for this happy future, by calculating the  
numerical factors that enter into the reconnection probability $P$.  To  
be maximally prepared, we consider not only the fundamental F-strings,  
but also D-strings and bound states of F and D strings.

In \S2 we give a primer on the properties of the F- and D-strings of  
the IIB string theory, which we hope will be broadly accessible.  We  
also briefly discuss more general string theories.
In \S3 we consider F--F collisions.  The basic method was developed in  
ref.~\cite{coll}, wrapping the F-strings on a large torus and using  
unitarity to relate the reconnection probability to a four-point tree  
amplitude.  Here we extend the result to the supersymmetric case.  In  
\S4 we consider F--D and F--$(p,q)$ collisions, using a similar  
unitarity method~\cite{break}.
We also show that the F--X reconnection probability, for all string  
types X, can be written in a universal form in terms of the long  
distance supergravity fields of \mbox{X}.
In \S5 we consider D--D collisions, using the approach first set out in  
ref.~\cite{bachas}.  These
are more complicated than the collisions of F-strings, but we are able  
to obtain useful approximate results in various regimes.
In \S6 we consider the collisions of vertices in the string network,  
including the disconnection process which is the inverse to that  
considered in \S4 and \S5.

The calculations in \S3 to \S6 are for toroidal compactification, where  
the strings are free to move in a flat higher-dimensional spacetime.   
In realistic compactifications one expects that all flat directions  
will be lifted, and the strings will sit near the minimum of a  
potential in the compact directions.  We consider this situation in  
\S7, and show that if the scale of the potential is somewhat less than  
the scale of the string tension, then it is possible to use the flat  
spacetime calculation in combination with an effective compactification  
volume arising from the quantum fluctuations of the string.  The  
effective volume depends only logarithmically on the ratio of scales,  
not on powers, but nevertheless is numerically important.  In \S8 we  
bring all of our results together for some representative models.  We  
find that $P$ in most cases is less than one, so that F- and D-strings  
can in principle be distinguished from gauge theory strings.  In some  
cases $P$ is as small as $10^{-3}$, which would have a large effect on  
the behavior of string networks.  The reconnection probability for  
strings of {\it different} types depends strongly on the details of the  
compactification and can either be rather large or essentially zero.   
In either case the effect on the string networks can be large, as we  
discuss in the conclusions.

\subsection*{Reconnection in field theory: a brief review}

Let us briefly review the situation with reconnection of field theory strings.  Magnetic flux tubes are classical gauge theory solitons.  If the gauge theory is perturbative then the string evolution is described by the classical field equations and is deterministic: for given incoming velocity and angle the strings either reconnect or they do not.  In adiabatic collisions they always reconnect, because this allows the flux (and the zero of the Higgs field) to take an energetically favorable shortcut.  Simulations of the classical equations show that this persists up to a center of mass velocity of around $0.9 c$~\cite{matz}, above which the strings pass through one another.\footnote{We know of no good analytic determination of this crossover velocity.  Ref.~\cite{hanhash} has constructed an analytic model of both soliton and D-string scattering.  This appears to have a lower crossover velocity; it would be interesting to understand the difference, but for now we assume that the simulations capture the details of the string interaction more completely.}  Cosmic string networks are moderately relativistic, $v \sim 0.6$-$0.7c$.  Only 1-2\% of collisions will reach $0.9c$ in the center of mass, so the reconnection probability $P$ is essentially one.

Of course, the simulations consider only the simplest field theory models.  Consider a classical field theory with a continuous global symmetry that is unbroken in vacuum but broken in the string core.  The string will have an additional collective coordinate, analogous to motion in a higher dimensional space, and two strings might avoid each other due to their separation in this coordinate~\cite{cargese}.  This was explored in ref.~\cite{hashtong}, where it was found that $P$ remains unity at least in the moduli space limit.  Ref.~\cite{hashtong} also observes that for a broken {\it discrete} symmetry there will be $N$ types of string with self-reconnection probability $P_{\rm s} = 1$ and nonself-reconnection probability $P_{\rm ns} <1$ (depending on the collision energy and barrier height).   Taking for simplicity $P_{\rm ns} = 0$, one gets an `average' $P$ equal to $1/N$.

This last model provides a nice illustration of the potential of cosmic string phenomenology.  For a single string type, two things happen as $P$ is reduced: (1) the number of long strings increases, most likely as $1/P$, so as to give the same long-string reconnection rate~\cite{pl1}; (2) the short-distance kinkiness of the strings increases, so that number of self-intersections increases by a factor $1/P$, so as to give the same loop production rate per long string (as required for the network to scale~\cite{ACK}).  With $P_{\rm s} = 1, P_{\rm ns} = 0$ one will have the first effect but not the second.  If we are fortunate enough to see the string network directly, either through lensing or its effect on the CMB, then it should be possible to distinguish this situation from a single species with $P = 1/N$ in short order.

Another possibility is gauge theory electric flux tubes~\cite{witten}, which would have a reconnection probability of order $1/N_c^2$.  Also, in a gauge theory with $\alpha_{\rm YM}  \sim 1$, or a string theory with $g_{\rm string} \sim 1$, the distinction between different kinds of objects disappears and it becomes difficult to identify distinctive signatures (of course the same is true of ordinary accelerator signatures as well).  However, as the example in the previous paragraph shows, the study of cosmic string networks has the potential to differentiate between seemingly similar microscopic models, even if not to resolve all degeneracies, and so we should not be too pessimistic.

\sect{A primer on F- and D-strings}

We will focus primarily on the IIB string theory, both because it  
provides the most well-developed string model of inflationary  
cosmology~\cite{KKLMMT} and because it has a potentially rich set of  
cosmic strings.  The fundamental IIB string has a tension that we will  
denote $\mu_{\rm F}$,
whose value might lie anywhere between the TeV scale and the Planck  
scale, though in the brane inflation models that provide much of our  
motivation one expects the narrower range
$10^{-12} < G \mu_{\rm F} < 10^{-6}$~[2-6].  The string in principle  
oscillates in all nine spatial dimensions, but as we will discuss in  
\S7 one expects the oscillations in the compact directions acquire  
nonzero world-sheet masses.  It also has neutral fermionic degrees of  
freedom, but these will be massive as well as a consequence of  
supersymmetry breaking.

Besides its tension, other important properties of a cosmic string are  
its couplings to axions and to Standard Model fields.  In the models of  
ref.~\cite{CMP}, all strings are non-axionic.  That is, they couple to  
massless potentials in ten-dimensions, but there are no light modes of  
these fields in the four-dimensional gauge theory.
More generally, all axions in string theory are expected to have  
instanton-generated potentials, so that axionic strings would be  
confined and uninteresting as cosmic strings~\cite{VE,witten}.  (It is  
conceivable that there are models where the instanton action is very  
small and there are interesting axionic strings.)  Also in the models  
of ref.~\cite{CMP}, stability of the cosmic strings requires in most  
cases that the strings be physically separated in the compact  
directions from Standard Model and other light fields, so they will be  
somewhat decoupled from these and in particular will not be  
superconducting.

The IIB string theory also has odd-dimensional D-branes~\cite{joed}, in  
particular D-strings. The D-string is much like the F-string except for  
its tension,
\begin{equation}
\mu_{\rm D} = \frac{\mu_{\rm F}}{g_{\rm s}}\  .
\end{equation}
The dimensionless string coupling $g_{\rm s}$ is determined by the  
expectation value of the dilaton field, $g_{\rm s} = e^{\Phi}$ (we are  
temporarily setting the RR scalar to zero, but will include it  
shortly).  If the dilaton varies in the compact directions, its value  
at the position of the string is the relevant one.  There is a special  
class of models in which $g_{\rm s}$ is related to observed parameters.  
  Namely, if the Standard Model fields live on D3-branes, then  
$\alpha_{\rm GUT} = \tilde g_{\rm s}$, where the tilde refers to the  
value of the dilaton at the position of the Standard Model D3-brane; if  
in addition the dilaton is constant as in IIB orientifold models, then  
$g_{\rm s} = \tilde g_{\rm s} = \alpha_{\rm GUT}$.  Unification in such  
models is necessarily nonstandard (see e.g.~\cite{unif}), but generally  
$\alpha_{\rm GUT} \sim 0.05$.  In F-theory compactifications on the  
other hand, the dilaton varies strongly over the compact space and  
there is no prediction for $g_{\rm s}$.  Our calculations use  
perturbation theory in $g_{\rm s}$ and so are quantitatively valid only  
if $g_{\rm s}$ is somewhat less than one.  Note that  if $g_{\rm s} >  
1$ we may switch to a dual description in which F and D strings are  
interchanged and $g'_{\rm s} = 1/g_{\rm s} < 1$.

Furthermore, $p$ F-strings and $q$ D-strings can bind to form a $(p,q)$  
string with tension
\begin{equation}
\mu_{(p,q)} =  \frac{\mu_{\rm F}}{g_{\rm s}} \sqrt{(p - Cq)^2 g_{\rm  
s}^2 + q^2}\ .
\label{genten}
\end{equation}
We have now included the expectation value of the RR scalar $C$.  This  
multiplet of strings was discovered through the $SL(2,Z)$ duality of  
the IIB string~\cite{pq}, and later explained in terms of FD bound  
states~\cite{bound}.
For relatively prime $p$ and $q$, the $(p,q)$ string has a nonzero  
binding tension with respect to any decomposition.  For $p = np'$ and  
$q=nq'$, the $(p,q)$ string is neutrally stable to splitting into $n$  
$(p', q')$ strings.  Supersymmetry breaking will lead to a weak  
potential between them, but this is likely to be negligible for most  
purposes.  The $n$ strings will in any case move on a common classical  
trajectory.

The integers $(p,q)$ can have either sign and are defined with respect  
to a specified orientation of the string; reversing the orientation  
sends $(p,q) \to (-p,-q)$.
Three $(p,q)$ strings can meet provided
\begin{equation}
p_1+p_2+p_3 = q_1 + q_2 + q_3 = 0\ ,
\end{equation}
where all strings are defined as pointing into the vertex.  The vertex  
is essentially massless,
so the angles at which the strings meet is fixed by the requirement  
that there be no force on the vertex.  In its rest frame this implies  
that the strings lie in a plane, and that the angle between strings $i$  
and $j$ is~\cite{angle}
\begin{equation}
\cos\hat\theta_{ij} = \frac{e_i \cdot e_j}{|e_i||e_j|}\ ,\quad e_i = (  
[p_i - Cq_i]g_{\rm s},q_i )\ .
\label{3ang}
\end{equation}

The string network evolves according to the Nambu-Goto action for each  
segment, with the constraint~(\ref{3ang}) where segments meet.  This  
constraint can also be incorporated into the dynamics by assigning a  
small mass to the junction and allowing it to respond to the tensions  
of the attached strings.

The network can also change discontinuously when two strings collide.
When $(p_1,q_1)$ and $(p_2,q_2)$ strings collide, they will pass  
through one another or reconnect in one of two ways as in  
figure~2.\footnote{Another possibility would be that the two strings  
would stick together at a four-string vertex.  Such higher vertices  
exist in some string models, where they arise from wrapped branes and  
are referred to as `baryons' because of their role in gauge/string  
duality.  However, they are rather massive and so unlikely to form in  
most cases.}
\begin{figure}[h]
\begin{center}
\leavevmode
\epsfbox{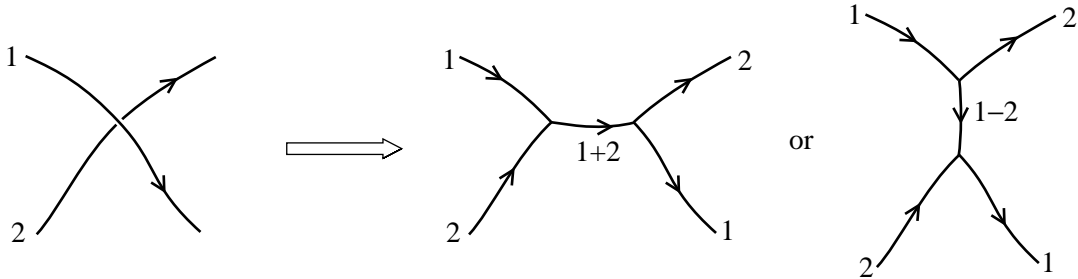}
\end{center}
\caption{Possible reconnections of $(p_1,q_1)$ and $(p_2,q_2)$ strings;
$i$ stands for $(p_i,q_i)$.}
\end{figure}
If $\theta$ is less than the angle $\hat\theta_{12}$ defined in  
eq.~(\ref{3ang}), it is energetically favorable to form a segment of  
$(p_1 + p_2, q_1 + q_2)$ string, and this is the presumed final state.   
If $\theta$ is greater than $\hat\theta_{12}$, the $(p_1 - p_2, q_1 -  
q_2)$ string is favored.
The behavior of string networks will also depend on what happens when  
two string vertices meet, the inverse of the process shown in figure~2.  
  We will take this subject up in \S6.

The $(p,q)$ strings are one-dimensional as seen either by a  
four-dimensional physicist or by a ten-dimensional physicist.   
Four-dimensional strings can also arise from higher-dimensional  
objects, $p$-branes\footnote{We use the standard terminology, but note  
that this is a different $p$ from the $(p,q)$ strings.} in which $p-1$  
dimensions are wrapped on part of the compact space and only one is  
extended in the visible directions.  The IIB string has odd-dimensional  
D-branes, for example.  Our results for D-strings (D1-branes) can be  
extended to these in a straightforward way. The IIB string also has an  
NS5-brane, which is a classical soliton like the ordinary cosmic  
string.  When inflation arises from a D3/anti-D3 system, as for example  
in the models~\cite{KKLMMT}, only one-dimensional branes are  
produced~\cite{tye2}.

The IIA theory has a fundamental string, even-dimensional D-branes and  
an NS5-brane, and the same considerations apply.  Similarly the type I  
theory has D1 and D5-branes (the fundamental type I string decays  
immediately by breakage on the space-filling D9), the heterotic theory  
has an F-string and an NS5-brane, and M-theory has M2- and M5-branes.   
The heterotic theory also has fundamental gauge fields, and associated  
with these there can be electric and magnetic flux tubes which may be  
interesting cosmic strings~\cite{witten}; the same applies to the low  
energy gauge theories that arise on branes.  Indeed, it seems that  
these should be regarded, roughly speaking, as dual to the non-BPS  
strings identified in ref.~\cite{CMP}.  In particular the magnetic flux  
tubes and D-strings both have large reconnection probabilities, while  
the electric flux tubes and F-strings have reconnection probabilities  
suppressed by $1/N_{\rm c}^2$ and $g_{\rm s}^2$ respectively.

\sect{F--F reconnection}

\subsection{Leading order}

We wish to calculate the probability for the process shown in figure~1.  
  This was done in ref.\ \cite{coll} for the bosonic string; we review  
the method and extend it to the supersymmetric case.

Locally near the intersection, the process is two infinite straight  
strings going to two infinite bent strings.  To make the process  
four-dimensional we compactify the
six transverse dimensions on a six-torus of volume $V_\perp$.
In order to use familiar vertex operator methods, we also wrap the long  
strings on a two-torus of lengths $l_{1,2}$ and angle $\theta$; at the  
end we take the two-torus volume to infinity.  The resulting process,  
shown in figure~3, is two unexcited winding strings going to an excited  
winding string with two kinks.
\begin{figure}[h]
\begin{center}
\leavevmode
\epsfbox{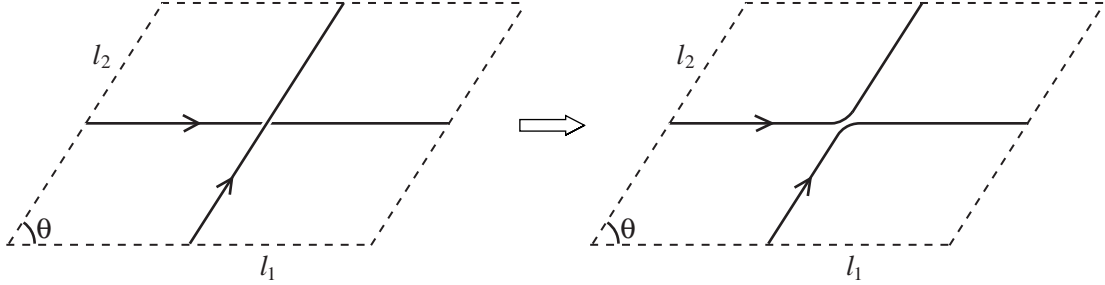}
\end{center}
\caption{F--F reconnection with strings wound on a torus.  This is a  
closed+closed$\,\to\,$closed transition.}
\end{figure}
The vertex operators for the two initial strings are much simpler than  
that for the final string.  Fortunately, since we are interested in the  
total interaction probability, we can square and sum over intermediate  
states.  By unitarity, this is related to the imaginary part of the  
tree level amplitude with four unexcited winding strings.

To simplify the supersymmetric calculation, we take the toroidal  
identifications to include a factor of $(-1)^{\bf F}$, where ${\bf F}$  
is the spacetime fermion number. That is, we are treating both  
directions on the torus as `temperature' directions.  This gives the  
opposite of the usual GSO projection, so the ground states are scalars.
Physically, adding a finite number of excitations to a long string  
cannot change the result.

The quantization conditions for the ground state strings of unit  
winding number are then
\begin{equation}
p_L^2 = p_R^2 = \frac2\ap\ ,\quad p_{L/R} = p \pm
\frac{L}{2\pi\ap}\ .
\end{equation}
The vertex operators in the $(-1,-1)$ and $(0,0)$ pictures are:
\begin{eqnarray}
   \V^{\,(-1,-1)} &=& \frac{\kappa}{2\pi\sqrt  
V}\,\NOrder{e^{-\phi-\tilde\phi + i  p_L  \cdot X+ i p_R \cdot\tilde  
X}}\ ,\nonumber\\
   \V^{\,(0,0)} &=& \frac{ \kappa}{2\pi
   \sqrt V} \frac{\ap}{2}  (\psi\cdot p_L)(\tilde\psi\cdot p_R)\,
   \NOrder{e^{ip_L  \cdot X+ i p_R \cdot\tilde X}}\ .
\end{eqnarray}
Here $V = {V_\perp l_1 l_2 \sin\theta}$ is the eight-dimensional  
compactification volume, and the factor of $V^{-1/2}$ is from the zero  
modes.

The appropriate vertex operator correlator is
\begin{equation}
   \A = \Expect{\V^{\,(0,0)}_1 \V^{\,(0,0)}_2
     \V^{\,(-1,-1)}_3 \V^{\,(-1,-1)}_4},
\end{equation}
where by the Riemann-Roch theorem there must be two vertex operators
in the $-1$ picture, and the position of the first three operators are
fixed to $z=0$, $z=1$ and $z\to\infty$ with $c$-ghosts.  Also, because
we use the optical theorem, we shall later set $p^3 = -p^1$ and $p^4 =
-p^2$.  Then
\begin{equation}
   \A =  (2\pi)^2\delta^{(2)}({\textstyle \sum_i} p_i) \frac{N_{\mathbbm  
S^2} \kappa^4}{(2\pi)^4 V^2}
   \Bigl(\frac\ap2 p_{L1}\cdot p_{L2}\Bigr)^2
   |z_4|^{\ap p_{L1}\cdot p_{L4}}
   |1-z_4|^{\ap p_{L2}\cdot p_{L4}}.
\end{equation}
The normalization of the path integral is crucial, and is $N_{\mathbbm  
S^2} = 32\pi^3V/\kappa^2\ap$ \cite{joev2}; the additional volume of  
compactification,
$V$ comes from the zero-mode integrals.  We have
used the fact that in this calculation there is never momentum in the
wound directions, so $p_{Li}\cdot p_{Lj} = p_{Ri}\cdot p_{Rj}$.
The positions of the first three operators are fixed; integrating over  
the fourth gives an invariant amplitude
\begin{equation}\label{virshap}
   \mathcal M = - \frac{4 \kappa^2 }{ V \ap }
    \frac{\GF{-\frac\ap4s}\GF{-\frac\ap4t}\GF{-\frac\ap4u}}
        {\GF{1+\frac\ap4s}\GF{1+\frac\ap4t}\GF{1+\frac\ap4u}},
\end{equation}
where $s,t$ and $u$ are the Mandelstam variables, constructed from  
either of $p_{Li}$ or $p_{Ri}$.

We construct the angled F-string pair wrapped on the torus, with one  
string
stationary and the other travelling toward it at velocity $v$, by  
setting
\begin{eqnarray}\nonumber
   p_1 &=&  
\biggl[\biggl(\frac{l_1}{2\pi\ap}\biggr)^2-\frac2\ap\biggr]^\hf
     (1,0,0,0,\mb0)\ ,\quad
   L_1 = {l_1}(0,1,0,0,\mb0),\\ \label{p2def}
   p_2 &=&  
\biggl[\biggl(\frac{l_2}{2\pi\ap}\biggr)^2-\frac2\ap\biggr]^\hf
     [1-v^2]^{-\hf}(1,0,0,v,\mb0)\ ,\quad
   L_2 = {l_2}(0,\cos\theta,\sin\theta,0,\mb0)\ .\qquad
\end{eqnarray}
For $l \gg \sqrt{\ap}$ with fixed small $t$ (corresponding to momentum  
transfer in the transverse directions) we are in the Regge region  
and\footnote{The imaginary part of the exact expression~(\ref{virshap})  
lies at discrete poles; this discreteness arises from the introduction  
of the two-torus.  In using Stirling's approximation we average these  
poles into a cut, as appropriate for the large-$l_i$ limit.}
\begin{equation}
   \mathcal M =  -
\frac{\kappa^2 }{ V } \frac{s^2}{t}
  (\ap s/4)^{\ap t/2} e^{-i \pi \ap t/4}\ . \label{regge}
\end{equation}
The normalization of the $t = 0$ pole agrees with graviton exchange  
calculated in an effective field theory, and the full form is  
determined by this normalization plus the Regge behavior.  By an  
extension of this observation we will be able to obtain the general  
F--X reconnection probability from field theory.

Inserting standard kinematic factors, the optical theorem gives
\begin{eqnarray}\nonumber
   P &=& \frac1{4 E_1 E_2 v}2\,\Im\mathcal M|_{t=0}\\
   &=& \frac{\kappa^2}{\ap \pi V_\perp} f(\theta,v)\ ,\quad  f(\theta,v)  
=
\frac{  (1-\cos\theta\sqrt{1-v^2})^2  }{8\sin\theta\, v\sqrt{1-v^2}}
\ .
   \label{FFProb}
\end{eqnarray}
Note that the factors of $l_i$ have cancelled out to give a finite $l_i  
\to \infty$ limit.  The result is the same as for the bosonic  
string~\cite{coll}, as we could have anticipated from the remark below  
eq.~(\ref{regge}).  That the center of mass frame was used in  
ref.~\cite{coll}, so our velocity $v$ is related to the velocity $v'$  
there by $v = 2v'/(1+v'^2)$.  Also, we have corrected an error in the  
sign of the $\cos\theta$ term; the numerator in $f(\theta,v)$ now goes  
to zero in the supersymmetric limit $\theta = v = 0$.

In terms of the dimensionless IIB string coupling $g_{\rm s}$,
\begin{equation}
P = \frac{g_{\rm s}^2 (2\pi)^6 \ap^3 }{V_\perp} f(\theta,v)
= {g_{\rm s}^2 } \frac{V_{\rm min}}{V_\perp} f(\theta,v)\ ,   
\label{FFProb2}
\end{equation}
where $V_{\rm min} = (4\pi^2 \ap)^3$ is the minimum volume of a  
six-torus in the sense of $T$-duality.  This is dimensionless, as it  
must be.  If instead we leave $d > 4$ dimensions noncompact, the  
strings can miss each other and the relevant quantity is a  
cross-section of dimension [length]$^{d-4}$.  The difference is all in  
the zero modes, and the cross section is still given by  
eq.~(\ref{FFProb2}), with $V_\perp$ the volume of the $(10-d)$-torus  
but $V_{\rm min}$ unchanged.

\subsection{Higher corrections}

It would be interesting to consider higher corrections to our result.   
These can come from higher loops in string perturbation theory and from  
processes in which additional closed strings are emitted during the  
reconnection process.\footnote{Processes in which additional winding  
strings are created, or the winding pair are deflected, involve a  
length of string of order $l$ and so are exponentially suppressed in  
$l$.  This is similar to the exponential suppression of high-energy  
fixed-angle scattering, with winding in place of momentum~\cite{CK}.}
For example, at order $g_{\rm s}^4$, there will be the one-loop  
correction to the reconnection process and also reconnection with the  
emission of one small closed string.  There is no reason to expect that  
such corrections are unusually large, but it would be useful to have  
some estimate of their magnitude.

To obtain the next order corrections will require some refinement of  
the simple unitary argument.  The imaginary part of the torus  
amplitude, representing the total interaction probability at order  
$g_{\rm s}^4$, contains the above processes but also other  
non-reconnection processes.  In particular, there is an $O(g_{\rm  
s}^2)$ amplitude for the strings to pass through one another but for  
some of their oscillators to become excited in the process, for example  
from the gravitational interaction between them, and this process  
actually dominates the interaction probability at $O(g_{\rm s}^4)$.   
Because each string produces a conic geometry, after they pass through  
one another they are no longer straight: each has a kink, its ends  
being misaligned by the deficit angle of order $O(g_{\rm s}^4)$.  This  
kink can be described as a coherent state, an eigenstate of the mode  
operators $\alpha(k), \tilde\alpha(k)$ (these are the continuum version  
of the usual $\alpha_n, \tilde\alpha_n$, taking the limit of an  
infinite string).  A linear kink in $X$ translates into $\alpha(k),  
\tilde\alpha(k) \propto 1/k$.  The norm of such a state is
\begin{equation}
\int_{1/l}^\infty \frac{dk}{k^3} \sim l^2\ , \label{irdiv}
\end{equation}
diverging in the large-volume limit.  This appears to agree with the  
string calculation: the $t \to 0$ behavior is given by the eikonal  
approximation~\cite{eikon}, and in gravity each loop brings in an  
additional power of $s \sim l^2$.

It is likely that one can deal with this problem, and still take  
advantage of unitarity, by separating the contributions of different  
channels to the imaginary part of the one-loop amplitude.  Obtaining a  
tractable form may be difficult; whereas the exponentially suppressed  
amplitudes are dominated by a saddle point in moduli  
space~\cite{mende}, the forward amplitude is not.  One expects that the  
physics of the reconnection process is local, and so the higher  
corrections will have a good $l\to\infty$ limit, unlike the  
IR-divergent process~(\ref{irdiv}).  It would be interesting to verify  
this.

\sect{F--$(p,q)$ reconnection}

When an F-string crosses a D-string it can break, leaving its endpoints  
attached to the D-string.
Taking both strings to be wound on a torus as in the previous case  
gives the process shown in figure~4.  This is a closed-to-open  
transition for the F-string.
\begin{figure}[h]
\begin{center}
\leavevmode
\epsfbox{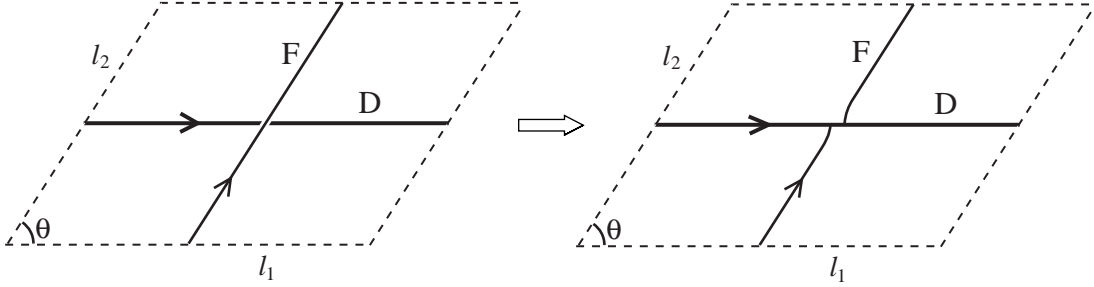}
\end{center}
\caption{F--D reconnection with strings wound on a torus.  This is a  
closed$\,\to\,$open transition.}
\end{figure}
Following ref.~\cite{break} we again use the optical theorem, obtaining  
the the total interaction probability from the imaginary part of the  
amplitude for two closed string vertex operators on the disk,
\begin{equation}
   \A = \Expect{\V^{\,(0,0)}_2    \V^{\,(-1,-1)}_4 }.
\end{equation}
  The vertex operators represent the winding F-string (and we have  
numbered them to match the previous section), while the D-string, which  
is stationary and oriented in the 1-direction, appears through the  
boundary conditions.  The difference from ref.~\cite{break} is that we  
are considering the superstring rather than the bosonic string, and  
D1-brane boundary conditions rather than the fully Neumann D25-brane  
boundary conditions.

Since it is straightforward to do so, we consider the general F--$(p,q)$
reconnection process.  The $(p,q)$ string gives a boundary with a  
$q$-valued Chan-Paton factor, while the $p$ bound F-strings appear as  
constant $U(1)$
electric flux on the D-branes.\footnote
{The bound state involves strongly coupled infrared dynamics for the  
$SU(q)$ degrees of freedom~\cite{bound}, but the process that we are  
considering takes place at the string scale and so is infrared-safe.}
The world-sheet CFT is free in such a background and the propagators  
are well known (see \emph{e.g.}~\cite{Seiberg:1999vs,garousi}),
\begin{eqnarray}
  \Expect{ X^{\mu}(z_1)X^{\nu}(z_2) }&=& - \frac\ap2 \eta^{\mu\nu} \ln  
(z_1-z_2)\ ,
  \quad
\langle \psi^{\mu}(z_1)\psi^{\nu}(z_2) \rangle =  
\frac{\eta^{\mu\nu}}{z_1-z_2}\ ,
  \nonumber\\
\langle X^{\mu}(z_1) \tilde X^{\nu}(\bar z_2) \rangle &=& - \frac\ap2  
G^{\mu\nu} \ln (z_1-\bar z_2)\ ,
   \quad
  \langle \psi^{\mu}(z_1) \tilde \psi^{\nu}(\bar z_2) \rangle =  
\frac{G^{\mu\nu} }{z_1-\bar z_2}\ . \label{xxdisk}
  \end{eqnarray}
Here
\begin{equation}
   G^{\mu\nu} = \left(\begin{array}{ccc}
     -\frac{1+f^2}{1-f^2}& \ -\frac{2f}{1-f^2}& 0\\[3pt]
     \frac{2f}{1-f^2}& \frac{1+f^2}{1-f^2}& 0\\[3pt]
     0& 0& -\mathbbm 1
   \end{array}\right)\ .
\end{equation}
This is the open string metric in the 0-1 plane, while the $-\mathbbm  
1$ reflects the Dirichlet boundary condition in the other directions.   
The parameter $f$ is related to the electric flux and the number of  
bound F- and D-strings by
\begin{equation}
  f = 2\pi\ap F_{01}\ , \quad  p = \frac{qf}{g_s\sqrt{1-f^2}}\ .   
\label{flux}
\end{equation}

Again we keep a finite momentum transfer before taking the limit, so  
that $p_4$ is $-p_2$ with a scattered velocity $\mb v'$ in the $3  
\ldots 9$ directions.  To leading order in $g_{\rm s}$ the D-string  
does not recoil and $| \mb v' | = v$.  The invariants relevant to the  
calculation are then given in terms of two variables $\sigma$ and $t$:
\begin{eqnarray}
\frac\ap2 p_{2L} \cdot G \cdot p_{2R} &=& \frac\ap2 p_{4L} \cdot G  
\cdot p_{4R} \equiv - \sigma \ ,\nonumber\\
\frac\ap2 p_{2L} \cdot G \cdot p_{4R} &=& \frac\ap2 p_{4L} \cdot G  
\cdot p_{2R} \equiv  \sigma
- \frac{\ap t}{4} \ ,\nonumber\\
\frac\ap2 p_{2L} \cdot p_{4L} &=& \frac\ap2  p_{2R} \cdot p_{4R} = -1 -  
\frac{\ap t}{4}\ .
\end{eqnarray}
We fix 3 of the 4 coordinates of the two vertex operators by
$z_2=i$ and $z_4 = ix$ with $x\in[0,1]$, and insert the corresponding  
$c$-ghosts. Evaluating the various factors in the expectation value  
gives
\begin{equation}
   \mathcal M = N_{\mathbbm D^2} \frac{\kappa^2\sigma 2^{-2\sigma}  
}{(2\pi)^2 V}  \int\limits_0^1 dx\,   
(1-x)^{-1-\alpha't/2}(1+x)^{1+2\sigma+\alpha' t/2}x^{-1-\sigma}\ .
\end{equation}
Making a change of variables to $x = (1-\sqrt{y})/(1+\sqrt{y})$ gives  
the standard representation of the beta function, and so~\cite{garousi}
\begin{eqnarray}
  \mathcal M &=& - N_{\mathbbm D^2} \frac{\kappa^2}{(2\pi)^2 V}
  \frac{\Gamma(-\frac\ap4t)\Gamma(1-\sigma)}{\Gamma(-\frac\ap4t-\sigma)}
  \nonumber\\
  &\stackrel{\rm Regge}{\to }&
  - N_{\mathbbm D^2} \frac{\kappa^2}{(2\pi)^2 V}  \frac{4}{\ap t}  
\sigma^{1+\ap t/4} e^{-i\pi t\ap/4}
    \ .
\end{eqnarray}

The normalization of the disc
partition function with these boundary conditions is $N_{\mathbbm D^2}
= 2 \pi^2 l_1 q\sqrt{1-f^2}/ 2\pi\ap g_s$.  This can be obtained from  
the standard disc
partition function normalization, $2\pi^2 V_9 \tau_9$, by $T$-duality,  
taking into account the Chan-Paton factors and the background fields.   
Note that each expression is $2\pi^2$ times the total Born-Infeld  
action for the D-branes.  One can check/verify this normalization by  
taking the extreme relativistic limit, where $\sigma \to 2 l_2  
/(1-f^2)(1-v^2)2\pi\alpha'$.  The $t=0$ pole is
\begin{equation}
\mathcal M  \to -\frac{\kappa^2}{V} \frac{2 l_1 \tau_{p,q} l_2^2}{t  
(2\pi\alpha')^2 (1-v^2)}\ .
\end{equation}
This has the same normalization as the pole in eq.~\ref{regge}, upon  
replacing
$s \to -2p_1 \cdot p_2 \to 2 m_{p,q} E_2$; note that we are in the  
$(p,q)$ string rest frame.  Also, there is a factor $1/2 m_{p,q}$  
because we are implicitly using canonical rather than relativistic  
normalization for the $(p,q)$ string.

Finally, the reconnection probability is
\begin{eqnarray}
P &=&  \frac1{2 E_2 v}2\,\Im\mathcal M|_{t=0} \nonumber\\
&=& g_{\rm s}^2 \frac{V_{\rm min}}{V_\perp} h_{p,q}(\theta,v)\ ,  
\nonumber\\
h_{p,q}(\theta,v) &=&
  \frac{q^2v^2+\left[  
g_sp-\cos\theta\sqrt{(1-v^2)(g_s^2p^2+q^2)}\right]^2}{8\sin\theta\;
     vg_s\sqrt{(1-v^2)(g_s^2p^2+q^2)}}  \label{FpqProb}
     \end{eqnarray}
To confirm this, we have also obtained it by a very different route:  
T-dualing the 1-direction and
boosting to give $q$ stationary D0 branes interacting with F-string  
winding states in motion.  Note that to leading order in perturbation  
theory, the $p$-dependence is important only if $p$ is of order  
$1/g_{\rm s}$.

The disk calculation of the F--$(p,q)$ probability~(\ref{FpqProb})  
requires at least one D1-brane, and so is not valid for $(p,q) =  
(1,0)$; in the latter case we have instead the sphere amplitude of \S3.  
  Nevertheless the probability~(\ref{FpqProb}) reduces to the earlier  
result~(\ref{FFProb2}) in this case, and so~(\ref{FpqProb}) is  
universal in $p$ and $q$.  This can be understood in a simple way.  We  
have already noted that the amplitudes can be normalized by a gravity  
calculation of the $1/t$ pole; taking this further, the full form of  
the $1/t$ pole can be obtained from a supergravity calculation.
The reconnection probability depends on the imaginary part as $t\to 0$.  
  This imaginary part comes in both calculations from
\begin{equation}
{\rm Im}\, e^{- i\pi\alpha' t/4} = -\sin(\pi \alpha' t/4)\ .  
\label{impart}
\end{equation}
The zero of the sine cancels the pole, giving a finite $t \to 0$ limit  
which is proportional to the supergravity amplitude.

We will not carry out the full supergravity calculation but we can  
check the $(p,q)$-dependence of some terms.  The $\cos^2 \theta$ term,  
which comes only from graviton exchange, is proportional to the tension  
of the $(p,q)$ string.  The $\cos \theta$ term, which comes only from  
$B_{\mu\nu}$ exchange, is proportional to the F-string charge $p$.   
(The angle-independent term is more complicated because both the  
graviton and the dilaton exchanges contribute.)  Also, at zero velocity  
the numerator vanishes when the angle between the strings is  
$\tan\theta = q/g_sp$,
which is the angle at which the strings are mutually BPS and there is  
no long-range force between them \cite{angle}.

We should emphasize that the imaginary part itself is not a  
supergravity effect.  It is analytic in $t$ and so local in spacetime.   
Rather, the connection is that the total cross section is given by the  
leading Regge trajectory at $t=0$, which is the supergravity amplitude.  
  The factor~(\ref{impart}), which arises from the continuation of Regge  
amplitude $(-s)^{\alpha' t/4}$ from Euclidean to Lorentzian momenta,  
provides the connection between the pole and the imaginary part.

Is this connection completely universal, so that we can immediately  
write down the general {$(p,q)$--$(p',q')$ result?  In the disk and  
sphere calculations, the factor~(\ref{impart}) has a common world-sheet  
origin.  In the Regge region, the in and out vertex operators $\V_2$  
and $\V_4$ are at small separation, and give a universal Regge form  
which is the same whether the rest of the world-sheet is a sphere with  
other vertex operators or a disk.  One might have expected this Regge  
behavior to be completely universal, but in the general  
$(p,q)$--$(p',q')$ interaction that we are about to consider, the  
world-sheet origin of the imaginary part is different, and the final  
result shows no sign of universality.

\sect{$(p,q)$--$(p',q')$ reconnection}

\subsection{Pair production}

When both strings have D1 charge, the leading interaction between them  
comes from annular world-sheets with one boundary on each.  This  
calculation was done for parallel D$p$-branes by Bachas~\cite{bachas};  
it has been extended and applied many times since, particularly in the  
small-velocity expansion relevant to matrix theory.

We start with the case of two $(p,q) = (0,1)$ D-strings, with one  
aligned at an angle
$\theta$ and travelling at speed $v \equiv \tanh(\pi\epsilon)$ relative  
to the other, and with impact parameter $y$.  It is straightforward to  
extend the results of ref.~\cite{bachas} to this case~\cite{lifschytz,  
BDL, arfaei, joev2},
\begin{eqnarray}\nonumber
   \M(y) &=& -\frac i2
   \int\limits_0^\infty \frac{dt}{t} e^{-{ty^2}/{2\pi\ap}}
   \left[\eta^6(it)\;\TF1{i\frac{\theta t}\pi}{it}
     \TF1{\epsilon t}{it}\right]^{-1}\\
   &&\qquad\qquad\qquad\times\left\{
   \sum\limits_{k=2}^4(-1)^{k-1}
   \TF{k}{0}{it}^2\TF{k}{i\frac{\theta t}\pi}{it}\TF{k}{\epsilon t}{it}
   \right\}.
   \label{VEff}
\end{eqnarray}
The total inelastic probability $P_{\rm pp}$ can be put in a simple  
universal form due to Schwinger~\cite{schwinger}.  This corresponds to  
production of at least one pair of stretched strings.  It is {\it not}  
the same as the reconnection probability, as we will explain in \S5.5.  
Summing over disconnected annuli gives
\begin{equation}
1 - P_{\rm pp}(y) = |e^{i\M(y)}|^2 = e^{-2\,\Im\M(y)}\ .
\end{equation}
The imaginary part arises from the poles of $\TF1{\epsilon t}{it}^{-1}$  
on the real $t$-axis, at $t = n/\epsilon$.  These are all traversed on  
the same side~\cite{bachas}:
\begin{equation}
{\rm Im}\,  \M(y) = \frac{1}{2} \sum_{n=1}^\infty \frac{1}{n} \Bigl[  
(-1)^{n+1} Z_B(n/\epsilon) + Z_F(n/\epsilon)
\Bigr]\ , \label{resid}
\end{equation}
where partition functions are
\begin{eqnarray}
Z_B(t) &\equiv& \sum_{{\rm bosons}\, i} e^{-2\pi \ap t m_i^2} =  
e^{-{ty^2}/{2\pi\ap}}
\frac{ \TF{3}{0}{it}^3\TF{3}{i\frac{\theta t}\pi}{it} -  
\TF{4}{0}{it}^3\TF{4}{i\frac{\theta t}\pi}{it}}   
{2\eta^9(it)\,i\TF1{i\frac{\theta t}\pi}{it}}
\ ,\nonumber\\
Z_F(t) &\equiv& \sum_{{\rm fermions}\, j} e^{-2\pi \ap t m_j^2} =  
e^{-{ty^2}/{2\pi\ap}}
\frac{  \TF{2}{0}{it}^3\TF{2}{i\frac{\theta t}\pi}{it} }   
{2\eta^9(it)\,i\TF1{i\frac{\theta t}\pi}{it}}\ .
\label{parts}
\end{eqnarray}
The residues sum up to give
\begin{equation}
1 - P_{\rm pp}(y) = \prod_{{\rm bosons}\, i}  (1+x_i)^{-1} \prod_{{\rm  
fermions}\, j} (1-x_j)\ , \label{1mp}
\end{equation}
where
$x = e^{-2\pi \alpha' m^2/ \epsilon}$
and $m$ is the mass of
the given stretched string state at minimum separation.

\subsection{Small velocity limit}

To get some understanding of this result, consider the limit $\epsilon  
\ll 1$.  Then
\begin{equation}
x = e^{2\pi \alpha' m^2/ \epsilon} \to \left\{ \begin{array}{c} 0 \ ,\  
\ m^2 > 0 \ , \\[3pt]
\infty\ ,\ \ m^2 < 0 \ . \end{array} \right.
\end{equation}
Thus $P_{\rm pp}(y) = 1$ if there is a tachyon in the spectrum, and  
$P_{\rm pp}(y) = 0$ otherwise.  This is the well-known fact that the  
annulus amplitude between
non-BPS configurations of branes diverges at the critical impact  
parameter where a tachyon first appears~\cite{banks}.  At larger impact  
parameters, the small-velocity process is adiabatic and elastic.  At  
smaller impact parameters, the tachyonic instability proceeds when the  
critical separation is reached.

The lightest string state in the present case is a boson with
\begin{equation}
m^2 = \frac{y^2}{(2\pi\alpha')^2} - \frac{\theta}{2\pi\alpha'}\ .
\end{equation}
This is tachyonic for $y^2 < 2\pi\alpha'\theta$~\cite{BDL}.  For  
strings at angles there is an obvious final state for the tachyonic  
decay, namely the reconnected strings~\cite{HT}; for some detailed  
studies of this process see refs.~\cite{angdecay}.

Note that the result is independent of the compactification volume,  
because the stretched strings are confined to the region near the  
intersection and have no zero modes.  Also, the D-strings are treated  
as having definite classical trajectories.
In the noncompact case this translates into a classical black-sphere  
cross section
\begin{equation}
\sigma = \int d^6 \mb y\, P_{\rm pp}(y) = (2\pi^2 \ap \theta)^3 \ .   
\label{cross}
\end{equation}
For toroidal compactification, taking the D-strings to have a constant  
wavefunction in the compact directions gives
\begin{equation}
P_{\rm pp} = \frac{(2\pi^2 \ap \theta)^3}{V_\perp} \ , \label{ddvol}
\end{equation}
where we assume that the $T^6$ is large enough to contain the black  
sphere without overlap.

We can now see that the supergravity argument does not extend to this  
case.  It would give the same as the small-velocity limit of the F--F  
probability, except for an additional factor of $g_{\rm s}^2$ from the  
greater tension of the D-string:
\begin{equation}
P_{\rm pp} \stackrel{?}{=} \frac{ (2\pi^2 \ap)^3 }{V_\perp}
\frac{  (1-\cos\theta)^2  }{v \sin\theta}
\ .
\end{equation}
In spite of the similarity of these expressions, they definitely differ  
by a factor of $v$, and there seems to be no way to relate them.  At  
the world-sheet level they have very different origins.  The Regge  
region for the F-string processes comes from small $z_{24}$,  
corresponding to a long cylinder in the $t$-channel.  The equivalent  
region for the annulus parameter $t$ (not to be confused with the  
Mandelstam $t$) is $t \ll 1$.
On the other hand, for small $\epsilon$ we see that the  
poles~(\ref{resid}) move to $t \gg 1$.

Of course, at $g_{\rm s} = 1$ F-strings and D-strings are identical  
under duality and so the reconnection probabilities should become  
equal.  This is not evident in the small-velocity limit, where the D--D  
interaction~(\ref{ddvol}) approaches a constant while the F--F  
result~(\ref{FFProb}) diverges as $1/v$.  Higher order effects must cut  
the latter off; perhaps some simple unitarization along the lines of  
the eikonal approximation can be used to estimate this.

\subsection{Ultrarelativistic velocities}

For $\epsilon \gg 1$, the poles in $t$ move to small values and so all  
string modes are significant in the sum~(\ref{resid}).  The asymptotics  
of the partition functions are given by a modular transformation,
\begin{equation}
Z_B(t) \sim Z_F(t) \simeq \frac{t^3}{4\sin\theta} e^{\pi/ t} + O(1) \ .
\end{equation}
The $n=1$ term in the sum~(\ref{resid}) dominates, giving
\begin{eqnarray}
2\,  {\rm Im}\,  \M(y) &\simeq& \frac{1}{2 \epsilon^3 \sin \theta}  
e^{\pi \epsilon -{ty^2}/{2\pi\ap}} \nonumber\\
&\simeq& K e^{-{ty^2}/{2\pi\ap}}\ ,\quad  K = \frac{8\pi^3}{\sin\theta  
\sqrt{1-v^2} [-\ln (1-v^2)]^3}
  \ .
\label{ultra}
\end{eqnarray}
We can now carry out the integral~(\ref{cross}), using the fact that  
$K$ is large:
\begin{eqnarray}
\sigma &=& \pi^3 \int_0^\infty dy \, y^5 \Bigl(1 - e^{- 2\,  {\rm Im}\,  
  \M(y) } \Bigr)
\nonumber\\
&\simeq& \frac{(2\pi^2 \alpha')^3}{6} {(\ln K)^3}  \nonumber\\
&\simeq& \frac{(2\pi^2 \alpha')^3 }{48}{[ -\ln (1-v^2)]^3}
\ .
\end{eqnarray}
As noted in ref.~\cite{bachas}, this is similar to the hard scattering  
of F-strings.

\subsection{Cosmic collisions}

In cosmic string networks, the velocities are moderately relativistic,  
so that a typical string collision will have $v \sim 0.7$ or $\epsilon  
\sim 0.3$~\cite{cosmicstrings}.\footnote{Of course, in an evolving  
network there will be a distribution of collision parameters.}
Surprisingly, this is not so different from the small velocity limit,  
in that only the lightest open strings are produced.  An excited string  
with $m^2 = 1/\alpha'$ has
\begin{equation}
x = e^{-2\pi\alpha' m^2/\epsilon} \lsim 10^{-9}\ .
\end{equation}
When $x$ is small, it is the same as the probability to create a pair  
of strings in the given state, which is therefore negligible.  For even  
higher states this probability decreases faster than the density of  
states increases, so only unexcited strings are produced with any  
probability.  Thus we can restrict the partition  
functions~(\ref{parts}) to those states whose mass goes to zero with  
$\theta$:
\begin{eqnarray}
Z_B(t) &\simeq& e^{-{ty^2}/{2\pi\ap}} \frac{6 + 2 \cosh 2\theta t}{2  
\sinh \theta t} =
e^{-{ty^2}/{2\pi\ap}} [ e^{\theta t} + 7 e^{-\theta t} + \ldots ]\ ,
\nonumber\\
Z_F(t) &\simeq&e^{-{ty^2}/{2\pi\ap}}  \frac{8 \cosh \theta t}{2 \sinh  
\theta t} =
e^{-{ty^2}/{2\pi\ap}} [ 4 + 8 e^{-2\theta t} + \ldots ] \ .  
\label{partexp}
\end{eqnarray}

As we will explain in \S7, the  mean value of $y^2$ in a realistic  
situation will be of order $g_{\rm s}$, and so we set it to zero in the  
spirit of this perturbative calculation.  In \S7 we will estimate  
corrections to this approximation, and we will find that they are  
likely to be quite substantial.  Nevertheless an understanding of the  
case $y^2 = 0$ is instructive.
A state contributing $e^{-k\theta}$ in $Z_{B,F}$ then has
\begin{equation}
x \simeq e^{- k \theta / \epsilon}\ ,
\end{equation}
and so
\begin{equation}
1 - P_{\rm pp}(0) = (1 + e^{\theta / \epsilon})^{-1} (1-1)^4 (1 +  
e^{-\theta/\epsilon})^{-7} (1 - e^{-2\theta/\epsilon})^8
\cdots\ .
\end{equation}
For angles of order one, $e^{- \theta / \epsilon} \lsim 0.04$ and the  
first two terms in the infinite product dominate:
\begin{equation}
1 - P_{\rm pp}(0) \simeq (1+e^{\theta / \epsilon})^{-1} (1-1)^4  \lsim  
0.04 \times 0^4\ .
\label{lowest}
\end{equation}
That is, the production probability is at least 0.96 for a pair of  
the lightest bosonic strings, and exactly 1 for a pair of each of the  
four lightest fermionic strings.   The reason for the exact zero in  
$1-P_{\rm pp}(0)$ is that the mass of the lowest fermionic states  
passes through zero and so there is a level crossing: the empty  
in-state becomes the filled out-state.

For smaller angles the tachyonic term becomes smaller but the higher  
terms in the series rapidly begin to contribute.  Numerically, the  
probability to produce a pair, aside from the four fermionic zero mode  
pairs, reaches a minimum around $0.95$ for $\theta/\epsilon \sim 2$.   
In realistic situations, the fermionic zero modes will be lifted  
somewhat by couplings to supersymmetry-breaking fluxes.  However, the  
net $P_{\rm pp}(0)$ will in most cases remain close to one, as long as  
the effective impact parameter $y$ does not become too large.

\subsection{Reconnection}

For $g_{\rm s} \ll 1$, the D-strings are much heavier than the  
F-strings.  Production of a single pair of F-strings will not cause the  
D-strings to reconnect.  Rather, the D-strings will pass through one  
another and separate, stretching the F-strings as they do so.  Thinking  
of the D-strings as open string solitons, one would expect that of  
order $g_{\rm s}^{-1}$ F-strings must be produced in order to produce a  
substantial change in the state of the D-strings.

We can make this estimate precise as follows.  After the collision the  
system is as shown in figure~5.
\begin{figure}[h]
\begin{center}
\leavevmode
\epsfbox{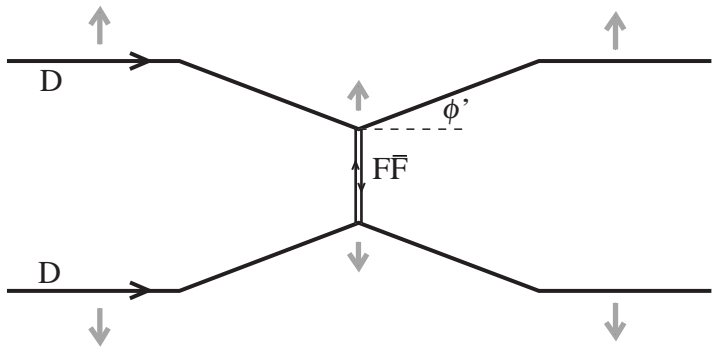}
\end{center}
\caption{D-strings after a collision that produces open F-string pairs.  
  The upper D-string is rotated by an angle $\theta$ relative to the  
lower, around the axis defined by the F-strings.  The angle $\phi'$ is  
the angle $\phi$ introduced in the text, boosted by the velocity of the  
vertex.}
\end{figure}
For $N$ F-string pairs, balance of forces implies that the angle  
between the F-strings and D-strings in the rest frame of the junction  
is $\half \pi + \phi$, where $\sin\phi = Ng_{\rm s}$.  The effect of  
the collision can only travel with the speed of light, so on each  
D-string there are two kinks traveling away from the point of the  
collision.  This picture only makes sense if $u > \tan\phi$, where $u =  
\tanh (\pi\epsilon/2)$ is the speed of each D-string in the  
center-of-mass frame.  If this is not satisfied the two junctions do  
not separate and the D-strings remain in contact (the vertices are at  
rest at the crossover velocity, so the boost of the junction angle does  
not enter).
For $u > \tan\phi$, where we have the situation in figure~5, the ends  
of the open strings are forced to remain localized and we expect that  
ends of opposite orientation will rapidly find one another and  
annihilate.  The two D-strings thus disconnect and continue onward,  
with some excitation.  For $u < \tan\phi$, we would expect that the  
D-strings will instead roll down to the reconnected configuration.   
Thus the condition for reconnection is $u < \tan\phi$, or equivalently
\begin{equation}
N > \frac{1}{g_{\rm s}} \sinh (\pi\epsilon/2)\ . \label{ddrecon}
\end{equation}

In order to apply this, we need to refine the earlier calculation,  
which just determined the probability that $N \geq 1$.  For the  
case~(\ref{lowest}) this is simple.  The tachyon pairs are produced in  
a squeezed state.  For a squeezed state of a single oscillator, if the  
probability of producing at least one pair is $p$, then the probability  
of producing at least $k$ pairs is
just $p^k$.  Here, $p \simeq 1 - e^{-\theta/\epsilon}$ and so
$p^k \simeq \exp(-k  e^{-\theta/\epsilon})$.  Counting the four  
fermionic pairs, the probability that the reconnection  
condition~(\ref{ddrecon}) is satisfied is
\begin{equation}
P =  \exp\Bigl( [ 4  - g_{\rm s}^{-1}  \sinh (\pi\epsilon/2)  ]   
e^{-\theta/\epsilon} \Bigr) \ .
\end{equation}
We see the somewhat surprising result that $P$ decreases as $g_{\rm s}  
\to 0$, because the reconnection condition~(\ref{ddrecon}) becomes more  
stringent while the probability of producing a given number of pairs is  
constant.  In fact, it falls as $e^{-O(1/g_{\rm s})}$ and so is  
nonperturbative, even though it was deduced from a perturbative  
calculation.  At asymptotically small couplings D--D reconnection is  
much less likely than F--F.\footnote{When $g_{\rm s}$, $v$, and  
$\theta$ all go to zero, the result depends on their ratio.}

This asymptotic suppression of $P$ does not set in until below the GUT value  
$g_{\rm s} \sim 0.05$.  For $g_{\rm s} \sim 0.05$ and $\epsilon \sim 0.3$, we
have $P \sim \exp(-  
6 e^{- \theta / 0.3} )$.  At $\theta \gsim 1$, $P$ is at least $0.8$; for  
$\theta \sim 0.6$ it falls to around 0.5 and then begins to rise again  
due to the higher states in the expansion~(\ref{partexp}).
Thus for this choice of parameters there is a range of small angles  
where D-strings will sometimes pass through one another, but this will  
have likely have a small effect on the network behavior.

\subsection{General $(p_1,q_1)$--$(p_2,q_2)$ interaction}

In the general case we have $q_1 q_2$ Chan-Paton states for the open  
strings, and an electric flux~(\ref{flux}) on each D-string from the  
dissolved F-strings.  The boundary conditions can be written  
as~\cite{Seiberg:1999vs}
\begin{eqnarray}
A^\mu{}_\nu X^{\nu\prime} (0) &=& B^\mu{}_\nu \dot X^\nu(0)\  
,\nonumber\\
C^\mu{}_\nu X^{\nu\prime}(\pi) &=& D^\mu{}_\nu \dot X^\nu(\pi)\ .
\end{eqnarray}
Here
\begin{equation}
A = \left[ \begin{array}{cccc}  1&0&0&0 \\ 0&1&0&0 \\ 0&0&0&0 \\ 0&0&0&0
\end{array} \right]\ , \quad B = B_1\ ,\quad C = AR\ ,\quad D= B_2 R\ ,
\end{equation}
defined in terms of
\begin{equation}
B_i = \left[ \begin{array}{cccc}  0&f_i&0&0 \\ f_i&0&0&0 \\ 0&0&1&0 \\  
0&0&0&1
\end{array} \right]\ ,\quad
R = \left[ \begin{array}{cccc}  \cosh \pi\epsilon &0&0& \sinh  
\pi\epsilon  \\ 0&\cos\theta&\sin\theta&0 \\ 0&-\sin\theta&\cos\theta&0  
\\  \sinh \pi\epsilon &0&0& \cosh \pi\epsilon
\end{array} \right]\ .
\end{equation}
Inserting a general linear combination of $e^{ i \omega(\tau+\sigma)}$  
and $e^{i \omega(\tau-\sigma)}$, one finds that $e^{2\pi i \omega}$  
must be an eigenvalue of
\begin{equation}
\Lambda = R^{-1} (A - B_2)^{-1} (A+B_2) R (A+B_1)^{-1} (A-B_1)\ .
\end{equation}
Noting that
\begin{equation}
(A+B_i)^{-1} (A-B_i) = \left[ \begin{array}{cccc}  \cosh 2\xi_i & -  
\sinh 2\xi_i &0&0 \\ -\sinh 2\xi_i &\cosh 2\xi_i&0&0 \\ 0&0&-1&0 \\  
0&0&0&-1
\end{array} \right]\ \in SO(1,3)\ ,
\end{equation}
where $f_i = \tanh \xi_i$, it follows that $\Lambda \in SO(1,3)$.  Any  
Lorentz transformation is conjugate to a boost times a commuting  
rotation, and so the eigenvalues take the same form
as in the D--D case
\begin{equation}
\omega = \pm{\tilde\theta}/\pi \ ,\ \pm i\tilde\epsilon\ ,
\end{equation}
in terms of an effective rotation angle $\tilde\theta$ and an effective  
rapidity $\pi\tilde\epsilon$.
For example, we can put these in the form (obtained using the $SL(2,C)$  
representation)
\begin{equation}
\cosh (\pi\tilde\epsilon + i \tilde\theta) = \cosh \xi_1 \cosh \xi_2  
\cosh (\pi\epsilon + i \theta)
- \sinh \xi_1 \sinh\xi_2\ ,
\end{equation}
whose real and imaginary parts determine $\tilde\theta$ and  
$\pi\tilde\epsilon$.  The full form is rather messy. One simple special  
case is a perpendicular $(p,q)$--D collision, $\theta = \pi/2$, $f_2 =  
0$, where $\tilde\theta=\pi/2$ and $\sinh\pi\tilde\epsilon =  
\cosh\xi_1\sinh \pi\epsilon$.

The discussion of D--D reconnection then extends directly to the  
general case.  Note that $(1-P)$ is raised to the power $q_1 q_2$ due  
to the Chan-Paton degeneracy.  The general conclusion, that  
reconnection almost always occurs unless $g_{\rm s}$ is very small,  
continues to hold.

There is one special circumstance that we should note.  Although they  
do not bind, $n$ fundamental strings can move together as though they  
were an $(n,0)$ string.  If these collide with another string, then in  
perturbation theory each will interact independently.
On the other hand, if we have $m$ coincident D-strings (or other  
$(p,q)$ strings with $q \neq 0$), and they cross a $(p',q')$ string  
with $q' \neq 0$, then the rule of thumb is that the tachyon decay will  
almost always take them to a new lower energy configuration, with a  
segment of $(p',q' \pm m)$ string, the sign depending on the angle.

\sect{Vertex interactions}

As the string network evolves, pairs of trilinear vertices will collide  
as in figure~6, and we need to determine the subsequent evolution.
\begin{figure}[h]
\begin{center}
\leavevmode
\epsfbox{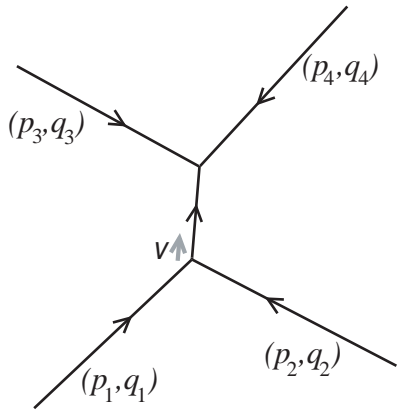}
\end{center}
\caption{Vertex collision: the lower vertex is moving toward the other  
with speed $v$.}
\end{figure}
In the figure, we begin with the supersymmetric configuration in which  
string $i$ is in the direction
$(p_i g_{\rm s}, q_i)$, and define a general configuration by rotating  
strings 3 and 4 by an angle $\psi$ around the string segment.
The simplest case is two F-strings ending on a $(p,q)$ string, so that
\begin{equation}
(p_i,q_i) = (p,q),\ (-1,0),\ (1,0),\ (-p,-q)\ ,\quad i = 1,2,3,4\ .
\end{equation}
The two F-strings have the same orientation, so the endpoints can  
annihilate and the F-string disconnect from the $(p,q)$ string.

We can obtain the probability for this by the same general strategy as  
for the previous F-string processes.  To set up the macroscopic open  
string states, we introduce two spectator
strings on which the other ends of the F-strings are fixed.  This is  
shown in figure~7; note that to lowest order the $(p,q)$ strings do not  
bend when a single F-string attaches.  These are at separation $R$ in  
the 2-3 plane, and we
take $R\to\infty$ at the end of the calculation to remove the
spectator strings and make the F-strings macroscopic.  As before we  
take the opposite of the usual GSO projection so as to get the simpler  
scalar ground state; this is equivalent to taking the spectators to be  
$(p,-q)$ strings.
\begin{figure}[h]
\begin{center}
\leavevmode
\epsfbox{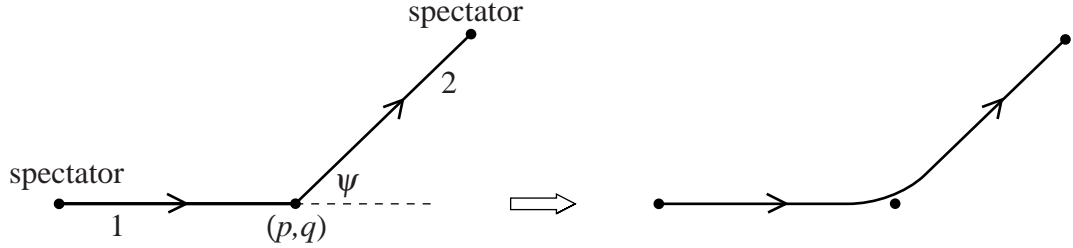}
\end{center}
\caption{Disconnection process as seen along the axis of the $(p,q)$  
string.  With two spectator $(p,q)$ strings it becomes an  
open+open$\,\to\,$open amplitude.}
\end{figure}
We again make
use of the optical theorem to obtain the contribution from all final
states.

The disconnection process is characterized by the relative velocity of  
the endpoints and the relative angle $\psi$ of the two strings  
(figure~6).  The open string vertex
operators for the strings stretched between the branes are
\begin{eqnarray}
   \V^{\,(-1)} &=& \lambda g_{\rm o}\, \NOrder{e^{-\phi} e^{ip_L  \cdot  
X+ i p_R \cdot\tilde X}}\ ,\nonumber\\
   \V^{\,(0)} &=& \lambda g_{\rm o}\,
   \sqrt{2\ap}(\psi\cdot p)\NOrder{e^{ip_L  \cdot X+ i p_R \cdot\tilde  
X}}\ .
\end{eqnarray}
We have added adjoint $U(q)$ Chan-Paton factors $\lambda$ to the  
F-string vertex operators.  Note also that there are no factors of  
$V_\perp$
since the open string wavefunctions are localized in the
extra dimensions by the D-strings.

We align the $(p,q)$-string along
the 1-direction, place the first F-string at rest aligned along the
2-direction, and the second with velocity along the $(p,q)$-string and
aligned at an angle $\psi$ in the 2-3 plane.  The momenta are therefore
\begin{eqnarray}\nonumber
  p_{1L,R} &=& \biggl[\biggl(\frac{R}{2\pi\ap}\biggr)^2-\frac{1}{2\ap}\biggr]^\hf [1-f^2]^{\hf}
    (1,0,0,0,\mb0) \pm  \frac{R}{2\pi\ap}(0,0,1,0,\mb0)\ ,\\ 
    p_{2L,R} &=& \biggl[\biggl(\frac{R}{2\pi\ap}\biggr)^2-\frac{1}{2\ap}\biggr]^\hf
    \biggl[ \frac{ 1-f^2}{1-v^2}\biggr]^{\hf} (1,v,0,0,\mb0)\pm
  \frac{R}{2\pi\ap}(0,0,\cos\psi,\sin\psi,\mb0)\ .\qquad
\end{eqnarray}
We have renumbered relative to figure~6: the initial and final  
F-strings are now 1,2 and 3,4 respectively.  The vertex operator for the
stationary string can be obtained, for example, by $T$-duality along the
1-direction, and then the other is obtained by a boost.
Using the contractions~(\ref{xxdisk}), each pair of vertex operators  
leads to a factor of $e^{2\ap p_i * p_j}$, where
\begin{equation}
p_i * p_j =  \frac{1}{4}( p_{iL} \cdot p_{jL} + p_{iR} \cdot p_{jR} +  
p_{iL} \cdot G \cdot  p_{jR} + p_{jL} \cdot G \cdot p_{iR} )\ .
\end{equation}
One can check the mass shell condition $p_i * p_i = 2/\alpha'$.

The amplitude is then
\begin{equation}
   \M = -\tilde N_{\mathbbm{D}^2} g_{\rm o}^4   
\Tr(\lambda^1\lambda^2\lambda^{2\dagger}\lambda^{1\dagger})
     \frac{\GF{-\ap s}\GF{-\ap t}}{\GF{1+\ap u}}\ ,
\end{equation}
where the Mandelstam variables are defined by $s = - (p_1 + p_2) * (p_1  
+ p_2)$ and so on.  The Chan-Paton trace is simply 1 (in all other  
channels it vanishes).  The path integral is normalized as in \S4,  
$\tilde N_{\mathbbm D^2} =  \pi\sqrt{1-f^2}/ \ap g_s$ (we are now  
treating the 1-direction with continuum normalization, and separating  
out the explicit Chan-Paton trace), while $\tilde N_{\mathbbm  
D^2}g_{\rm o}^2 =  1/\ap$ holds in general.  Using these normalizations  
and taking the imaginary part as in earlier calculations,
\begin{equation}
{\rm Im}\, \M = \frac{g_{\rm s} s}{\sqrt{1-f^2}}\ .
\end{equation}
With the usual kinematic factors, the disconnection probability is
\begin{equation}
P = \frac{2 {\rm Im}\, \M}{2E_1 2 E_2 v} = \frac{g_{\rm s}}{v} \frac{1  
 - \sqrt{1-v^2} \cos\psi}
{ (1 - f^2 )^{3/2} }\ .
\end{equation}

When three or four of the strings carry D-string charge, there is no  
simple CFT description of the system.  However, we can resort to the  
rule of thumb that in this situation the open string tachyons will  
almost always take the strings to their lowest energy state.  We must  
still determine which of strings 3 and 4 string 1 will join onto.  In  
fact, the supersymmetric configuration shown is neutrally stable in  
both directions, but any nonzero $\psi$ increases $\theta_{13}$ and  
decreases $\theta_{14}$ so that the latter reconnection is favored.  In  
particular, when $(p_1,q_1) = -(p_4,q_4)$, the strings disconnect.

\sect{Compactification effects}

The string reconnection  
probabilities~(\ref{FFProb2},\,\ref{FpqProb},\,\ref{ddvol}) all depend  
on the compactification volume as $V_{\rm min}/V_{\perp}$, reflecting  
the fact that the strings have to come roughly within a string radius  
in order to interact~\cite{tye3,tye4,DvV}.  It is therefore essential  
to determine the effective value of $V_{\perp}$.

Naively it would seem that one could obtain very small values of $P$ in  
models with large compact dimensions.  However, from the point of view  
of the world-sheet field theory, the position of the string in the  
compact dimensions is a scalar field, which is not protected by any  
symmetry.  One therefore expects that at some scale this modulus will  
be fixed, like the compactification moduli.  That is, there is an  
effective potential which localizes the string.
Moreover, the behavior of scalar fields in 1+1 dimensions implies that  
the effective volume over which the string wavefunction spreads depends  
only
logarithmically on the mass scale of the moduli --- as the cube of the  
logarithm, to be precise~\cite{CMP}.  As a result, $P$ can be  
suppressed somewhat, but not by many orders of magnitude.

\subsection{Generalities}

We obtain the effective action for a $(p,q)$ string moving in a general  
warped string metric
\begin{equation}
\label{metric}
ds^2 = H^{-1/2}(Y) \eta_{\mu \nu} dX^\mu dX^\nu+H^{1/2}(Y) g_{ij}(Y)  
dY^i dY^j \ ;
\end{equation}
we also allow the dilaton $\Phi(Y)$ to depend on the transverse  
coordinates.
We use $X^\mu$ for the noncompact dimensions and $Y^i$ for the  
transverse dimensions.
The relevant terms in the world-sheet action  for a $(p,q)$ string are
\begin{equation}
S = -\frac{ 1 }{2\pi\ap} \int d^2\sigma\,\nu(-\det h_{ab})^{1/2}\ ,
\end{equation}
where
\begin{equation}
\nu = (p^2 + q^2 e^{-2\Phi(Y)})^{1/2}\ ,\quad h_{ab} = H^{-1/2} (Y)  
\eta_{\mu \nu} \partial_a X^\mu \partial_b X^\nu + H^{1/2}(Y) g_{ij}(Y)  
\partial_a Y^i \partial_b Y^j
\ .
\end{equation}

To obtain first the potential, we insert a static configuration $X^0 =  
\sigma^0$, $X^1 = \sigma^1$,
$Y^i = {\rm constant}$.  The action then reduces to a potential
\begin{equation}
V(Y) = -{\cal L} = \frac{ \nu(Y) }{2\pi\ap H^{1/2}(Y)}\ . \label{potent}
\end{equation}
Compactifications with string tensions below the Planck scale generally  
have branes and fluxes that produce a nontrivial warp factor and/or dilaton, so  
that the potential~(\ref{potent}) depends nontrivially on the compact  
dimensions.  The strings will then sit near the minimum of the  
potential.  For strings that are supersymmetric with respect to all the  
branes the classical potential can cancel~\cite{tye3}, and it will then  
be necessary to go to higher order or even to nonperturbative physics  
to find the leading effect.

Notice that if the dilaton is nontrivial then the position of the  
minimum will depend on $p$ and $q$.  For the strongly warped  
geometries~\cite{KKLMMT} which are our main focus, the variation of the  
dilaton is negligible.
However, it should be noted that this effect has the possibility in  
principle to localize the different $(p,q)$ strings far enough apart  
that they will evolve as essentially independent networks.  Roughly  
speaking they must be separated by more than a string length for this  
to happen; we will make a few further comments below.

To understand the fluctuations around the minima we will need to expand  
the action to second order in the $Y^i$.  We choose coordinates such  
that the minimum is at $Y=0$ and $H^{1/2}(0) g_{ij}(0) = \delta_{ij}$.  Then
\begin{equation}
S \approx - \int d^2\sigma\, \biggl\{
V(0) + \frac{1}{2} \partial_i \partial_j V (0) Y^i Y^j +  
\frac{\nu(0)}{4\pi\alpha'} \partial_a Y^i \partial^a Y^i
\biggr\} \ .
\end{equation}
Notice that the $Y^i$ are now massive world-sheet fields.
We are interested in the average spread of the $Y^i$.  For a single  
scalar field with action
\begin{equation}
S = -\frac{Z}{2} \int d^2\sigma\,\Bigl (\partial_a \phi \partial^a \phi  
+ m^2 \phi^2 \Bigl)
  \ ,
\end{equation}
a Feynman diagram calculation gives
\begin{eqnarray}
\Expect{\phi^2(0)} &=& \frac{1}{Z} \int^\Lambda \frac{d^2 k}{(2\pi)^2}  
\frac{1}{k^2 + m^2}
\nonumber\\
&=& \frac{1}{4\pi Z} \ln  \frac{\Lambda^2 + m^2}{m^2} \ .
\end{eqnarray}
In our case, the UV cutoff is the string scale.  This is $\Lambda^2  
\sim1/\alpha'$ as seen by a ten-dimensional observer, but in  
four-dimensional units this is redshifted to $\Lambda^2 \sim
1/\alpha' H^{1/2}(0)$.  Rotating coordinates to make $\partial_i  
\partial_j V (0)$ diagonal, we have
\begin{equation}
\Expect{Y^i Y^i} = \frac{\alpha'}{2\nu(0)} \omega_i \ ,\quad \omega_i =  
  \ln \biggl[1 + \frac{ \nu(0)}{2\pi \ap^2 H^{1/2}(0) V_{,ii}(0) }\biggr]  
\qquad{\rm (no\ sum\ on\ }i) \ . \label{yy}
\end{equation}

 From this we learn that the fluctuations of the string in the  
transverse dimensions scale only logarithmically as we lower the scale  
of the potential that localizes the string; the linear scale of the  
fluctuations goes as the square root of the logarithm, and the volume  
goes as the cube of the logarithm.   The fluctuation~(\ref{yy}) is  
proportional to
\begin{equation}
\nu(0)^{-1} = \frac{g_{\rm s}}{(p^2 g_{\rm s}^{2} + q^2 )^{1/2}}\ ,
\end{equation}
so for strings with D-brane charge it vanishes to leading order in  
perturbation theory.

The calculation that we have done is meaningful only when  
$\Lambda^2/m^2$ is large, meaning that $V_{,ii}$ is small in string  
units.
This corresponds to the geometry varying slowly on the string  
scale.\footnote{Also in this limit the effective quartic coupling is  
small, so the only-loop calculation that we have done is valid.}
In this case we would expect to be able to combine the flat spacetime  
calculation that we have previously done with an effective wavefunction  
for the string calculated as above.  When $V_{,ii}$ is of order one in  
string units, so that there is no separation of scales, there is no way  
to use the flat spacetime calculation.  It is then necessary to do a  
full perturbative string calculation in curved space, and for the  
present we can only guess at the magnitude of $P$.
It is not clear whether there is any physical situation in which  
$V_{,ii}$ becomes much greater than one in string units, but if there is it
will  
require some complementary method of calculation.

\subsection{Effect on reconnection}

The effective value of $1/V_{\perp}$ is the density $\rho$ of the  
wavefunction where the strings coincide. For example, the effective  
density for an F--D collision is obtained from the field theory  
fluctuations for an F-string relative to a fixed center (since the  
fluctuations of the D-string are much smaller):
\begin{eqnarray}
V_{\rm min}\rho_{\rm FD} &=& V_{\rm min}\Expect{\delta^6 (Y)}  
\nonumber\\
  &=& V_{\rm min} \int  \frac{d^6 l}{(2\pi)^6}\, \Expect{ e^{i l \cdot  
Y} } \nonumber\\
  &=& V_{\rm min} \int  \frac{ d^6 l}{(2\pi)^6}\,  e^{-l_i l_j  
\Expect{Y^i Y^j} / 2    } \nonumber\\
  &=& \frac{ (4 \pi )^3}{ \prod_i \omega_i^{1/2} }  \ .
\end{eqnarray}
This has the expected logarithmic behavior, with the coefficient of the  
logarithm now determined.  This assumes that the F- and D-strings are  
localized at the same point.  If the F-string potential is minimized at  
$Y=0$ and the D-string potential at $Y = Y_{\rm D}$ then
\begin{eqnarray}
V_{\rm min}\rho_{\rm FD} &=& V_{\rm min}\Expect{\delta^6 (Y - Y_{\rm  
D})} \nonumber\\
  &=& \frac{ (4 \pi )^3}{ \prod_i \omega_i^{1/2} }
  \exp\biggl[ - \sum_i \frac { Y_{\rm D}^i Y_{\rm D}^i  }{ \alpha'  
\omega_i } \biggr] \ .
  \label{fdsupp}
\end{eqnarray}
This illustrates the expected large suppression when the separation of  
the minima is larger than the string scale.
For an F--F collision one has separate fields $Y$ and $Y'$, and so
\begin{eqnarray}
V_{\rm min} \rho_{\rm FF} &=& V_{\rm min} \Expect{\delta^6 (Y - Y')}  
\nonumber\\
  &=&  \frac{ (2 \pi )^3}{ \prod_i \omega_i^{1/2} }  \ . \label{rhoff}
\end{eqnarray}
Since the F-strings are of the same type their minima are coincident.

For the general $(p,q)$--$(p',q')$ collision the effect of a nonzero  
impact parameter $y$ is to replace
\begin{equation}
x \to xe^{-y^2/2\pi\alpha'\epsilon}
\end{equation}
in the general interaction probability~(\ref{1mp}).   Let us apply this  
in the case~(\ref{lowest}) that only the bosonic tachyon and the  
massless fermions are important, as holds over most of the parameter  
space.  Then
\begin{equation}
1 - P_{\rm pp}(0) \simeq  \frac{ (1-e^{  -y^2/2\pi\alpha' \epsilon})^4  
}{ 1+e^{\theta/\epsilon}
e^{ -y^2/2\pi\alpha'  \epsilon} }  \label{impact}
\ .
\end{equation}
Again we see that for strings in different minima, the interaction  
probability falls rapidly for separations large compared to the string  
scale.  For strings in the same minima, the quantum  
fluctuations~(\ref{yy}) imply that we must average the  
result~(\ref{impact}) over a gaussian wavefunction of this width.

A typical value of the correction factor is
\begin{equation}
e^{  -y^2/2\pi\alpha' \epsilon} \sim \exp\biggl[ -\frac{\sum_i   
\omega_i }{2\pi\nu(0)\epsilon} \biggr]\ ; \label{supp}
\end{equation}
we have assumed strings of the same type, so summing fluctuations as in  
eq.~(\ref{rhoff}) contributes a factor of 2 to the exponent.
For D-strings, $\nu^{-1}(0) = g_{\rm s}$, so this is formally higher  
order in perturbation theory.
However, the fluctuations in the different directions add to give an  
effective factor of 6, and so for the typical $\epsilon \sim 0.3$ the  
exponent can be of order one if the logarithm $\omega_i$ is large.  This would
lift the  
suppression due to the fermion zero modes.  The probability to produce  
tachyon string modes remains large until the suppression  
factor~(\ref{supp}) approaches $e^{-\theta/\epsilon}$, but if the scale  
of $V_{,ii}$ is low this can be the case, at least for some range of  
angles.  Thus there is the possibility that D-strings can pass through  
one another without reconnecting.

When the typical value of the exponent~(\ref{supp}) becomes large, the  
dominant contribution to the reconnection probability comes from those  
collisions that happen to occur at small impact parameter, near the  
center of the gaussian distribution.  In this case the reconnection  
probability is given by the ten-dimensional cross-section, which is  
given to reasonable approximation by the low-velocity  
limit~(\ref{cross}), times the peak density
\begin{equation}
\rho_{\rm DD} = \rho_{\rm FF}|_{\omega_i \to g_{\rm s} \omega_i} \ ;
\end{equation}
the factor of $g_{\rm s}$ reflecting the smaller fluctuations of the  
heavier D-string.

\subsection{Model parameters}

\subsubsection{The \Klmt model}

In the \klmt model~\cite{KKLMMT}, inflation takes place in a highly  
warped throat whose local geometry is given by the Klebanov-Strassler  
solution~\cite{KS}.  The warp factor also produces a potential well for  
the transverse coordinates of the string.  The geometry near the base  
of this solution is locally $\mathbb R^3 \times S^3$,
\begin{equation}
g_{ij}(Y) dY^i dY^j = dr^2 +  { r}^2 d\Omega_2^2+R_3^2 d\Omega_3^2\ ,
\end{equation}
where $R_3^2 = b g_{\rm s} M \alpha'$; $M$ is an integer characterizing  
the number of flux units.  The warp factor near the origin depends on  
the radial coordinate of $R^3$,
\begin{equation}
H(Y) = H(0)\biggl( 1 - \frac{b'r^2}{{ g_{\rm s} M \ap}} \biggr) \ .
\end{equation}
The energy scale of inflation in this model is of order $10^{-4}$ in  
Planck units, so $H^{-1/4}(0) \sim 10^{-4}$.
The constants $b \approx b'  \approx 0.93$~\cite{HKO} will be treated  
as 1 for our purposes.  The dilaton in this solution is constant.   
There are also three- and five-form fluxes in the compact directions,  
but these do not enter into the string action.  The product $g_{\rm s}  
M$ must be somewhat greater than one in order for the supergravity  
approximation to be valid.

This solution has the special property that the warp factor has its  
minimum not at a point but on the entire three-sphere at $r = 0$.  We  
first consider this geometry as it stands, and then consider  
corrections.  The effective $V_\perp$ is given by combining the volume  
of the $S^3$ with the quantum fluctuations on the $R^3$,
\begin{equation}
V_{\rm min} \rho_{\rm FF} \approx  \frac{ 4 \pi }{ (g_{\rm s} M )^{3/2}  
}
\frac{ (2 \pi )^{3/2}} {  \ln^{3/2}(1 +  g_{\rm s} M) }  \ .   
\label{s3vol}
\end{equation}
Also, $\rho_{\rm FD} = 2^{3/2} \rho_{\rm FF}$.

The fact that the potential is constant on the $S^3$ reflects an $SU(2)  
\times SU(2)$ symmetry of the Klebanov-Strassler solution.  This local  
geometry is part of a larger Calabi-Yau solution, which can have no  
isometries.  The breaking of the symmetry will generate an effective  
potential along the $S^3$, which will localize the strings.
The warp factor $H^{-1/4}(0)$ is a measure of the size of the tip of  
the Klebanov-Strassler throat in terms of the underlying Calabi-Yau  
geometry~\cite{GKP}.  It therefore governs the extent to which the  
throat feels the curvature of the geometry, and so size of the  
$SU(2)\times SU(2)$ breaking and the size of the potential.  We will  
assume that the curvature of the Calabi-Yau manifold will have an  
effect on the throat geometry of order the warp factor squared,  
$H^{-1/2}(0) \sim 10^{-8}$, relative to the other scales in the throat,  
so that $\omega_i \sim \ln H^{1/2}(0)$ in the $S^3$  
directions.\footnote{A more complete analysis might give a different  
power of $H(0)$ inside the logarithm. This will affect some of the  
numerical estimates, but not the overall logic.}
Then
\begin{equation}
V_{\rm min} \rho_{\rm FF} =  \frac{ (2 \pi )^3}{ \ln^{3/2}(H^{1/2}(0))  
\ln^{3/2}(1 +  g_{\rm s} M) }  \ .  \label{s3fluct}
  \end{equation}
In this case $\rho_{\rm FD}$ contains an additional factor of 8, but it  
may easily be the case that the F and D strings are localized at  
different points of the $S^3$, leading to the additional  
suppression~(\ref{fdsupp}).

The reader will notice that the density~(\ref{s3fluct}) might be less  
than the density~(\ref{s3vol}), depending on the parameters.  This is  
not a logical possibility.  What is happening is that the fluctuations  
begin to fill out the whole $S^3$; thus, we should always use whichever  
of the densities (\ref{s3fluct},\,\ref{s3vol}) is {\it greater}.  In  
the fluctuation calculation, the log of the ratio of scales times the  
world-sheet coupling $1/g_{\rm s}M$ is becoming large, and we must use  
the renormalization group to improve the calculation.

For D--D collisions, the relevant quantity is
\begin{equation}
\frac{y^2}{2\pi\alpha'  \epsilon}  \sim  \inf\biggl[ \frac{g_{\rm  
s}M}{2\pi\epsilon}, \frac{3g_{\rm s} \ln H(0)}{8\pi\epsilon } \biggr]\  
, \label{ddimp}
\end{equation}
depending on whether the quantum fluctuations fill out the $S^3$.
Again, the fermion zero modes are lifted to the extent that this is  
nonzero, and the tachyonic modes are not excited for collisions at  
angles less than $\epsilon \times(\ref{ddimp})$.

\subsubsection{Large dimension models}

Now let us consider models in which $n$ dimensions have  
periodicity~$2\pi R$ and $6-n$ have
the minimum periodicity $2\pi\sqrt\ap$~\cite{large}.  The four- and  
ten-dimensional gravitational couplings are
\begin{equation}
\kappa_4^2 = \kappa^2 (2\pi R)^{-n} (2\pi \sqrt\ap)^{n-6}\ .
\end{equation}
It is convenient to rewrite this as
\begin{equation}
G \mu_{\rm D} = \frac{g_{\rm s}}{16\pi} \biggl( \frac{\ap}{R^2}  
\biggr)^{n/2}
\end{equation}
  We will assume that the effects that fix the moduli and break  
supersymmetry produce modulations of the warp factor and/or the dilaton  
by a factor of order $\delta \lsim 1$, as opposed to the large warping  
of the \klmt model.  Then for F-strings, $V_{,ii} \sim \delta /2\pi  
\alpha' R^{2}$ in the large directions, and so
  \begin{eqnarray}
  \omega_{\rm large} &\sim& \ln \biggl( \frac{R^2}{\alpha'   
\delta}\biggr)
  \nonumber\\
  &\sim& \frac{2}{n}\ln\biggl( \frac{g_{\rm s}}{16\pi  G \mu_{\rm D}  
\delta}\biggr)
\ ,\nonumber\\
V_{\rm\perp} \rho_{\rm FF} &=& (2\pi/ \omega_{\rm large})^{n/2} \ .  
\label{fflarge}
\end{eqnarray}
In $\rho_{\rm FD}$ there is an additional $2^{n/2}$ but the likelihood  
of a suppression factor from separated minima.  For D--D collisions,  
the relevant quantity is
\begin{equation}
\frac{y^2}{2\pi\alpha'\epsilon  }  \sim \frac{ g_{\rm s}   
}{\pi\epsilon} \ln\biggl( \frac{g_{\rm s}}{16\pi  G \mu_{\rm D}  
\delta}\biggr)\ . \label{ddimplarge}
\end{equation}

\sect{Final results}

Having assembled all of the relevant calculations, it is interesting  
now to insert some typical parameter values and obtain estimates for  
$P$.  It is premature to take the details of these models too  
seriously, but it is important to get an idea of the range of  
possibilities --- both as a guide for the network simulations needed to  
estimate signals, and also to get some sense of the extent to which we  
might be able to probe stringy physics by measuring the various  
intercommutation probabilities.

Consider first F--F reconnection.  The reconnection probability is the  
earlier result~(\ref{FFProb2}) with one of  
eq.~(\ref{s3vol},\,\ref{s3fluct},\,\ref{fflarge}) in place of $V_{\rm  
min}/V_\perp$.  We will denote these cases respectively as (A) \klmt  
averaged over the $S^3$, (B) \klmt with fluctuations that do not fill  
out the $S^3$, and (C) large dimensions.  The function $f(v,\theta)$ is  
roughly 0.5 when averaged over angles and velocities.\footnote{We have  
noted at the end of \S5.2 that at small velocities unitarization  
effects must reduce $P_{\rm FF}$.  It is possible that these have some  
effect  at the velocities relevant for cosmic strings.}
   Then in the three cases
\begin{eqnarray}
P_{\rm FF}{\rm (A)} &\sim& \frac{100 g_{\rm s}^{1/2}}{M^{3/2}  
\ln^{3/2}(1+g_{\rm s} M)}\ ,
\nonumber\\
P_{\rm FF}{\rm (B)} &\sim& \frac{1.5 g_{\rm s}^{2}}{ \ln^{3/2}(1+g_{\rm  
s} M)}\ ,
\nonumber\\
P_{\rm FF}{\rm (C)} &\sim& 0.5 g_{\rm s}^{2}
\biggl[ \frac{\pi n}{\ln(g_{\rm s} / 16\pi  G \mu_{\rm D} \delta) }  
\biggr]^{n/2}
\end{eqnarray}

In case A, $P$ depends only weakly on $g_{\rm s}$ but strongly on $M$.   
The stability of these models requires $M$ to be at least 12, perhaps  
somewhat larger~\cite{KKLT}.  For $M = 20$, gives $P \sim 0.25$ over  
the range of $g_{\rm s}$ between 1 and the GUT value 0.05, while for $M  
= 100$, $P \sim 0.01$.  Decreasing the value of $P$ increases the total  
density of string in the network roughly as $P^{-1}$, because the  
reconnection process is needed for long strings to decay.  Thus, one  
might say that for $M = 20$, it would be possible with precise  
observations and simulations to distinguish these strings from the  
gauge theory strings that have $P=1$, while for $M=100$ there would be  
a substantial enhancement of the string density.  In case B, we should  
note that the supergravity approximation used in \klmt requires $g_{\rm  
s} M$ to be greater than one, but its logarithm need not be especially  
large.  In this case, one has $P \lsim  g_{\rm s}^2$, so $P$  
approaches one in the `worst case' that $g_{\rm s} \sim 1$, and it is  
much less than one for $g_{\rm s}$ near the GUT value.
Notice, following the discussion after eq.~(\ref{s3fluct}), that at any  
given point in \klmt parameter space one should use whichever of  A and  
B is {\it larger}.
In the case C, the scale $G\mu_{\rm D}$ is likely to lie between  
$10^{-6}$ and $10^{-12}$~\cite{tye2,tye3}.  Then for $\delta = 1$ and $n=2$,
the  
factor in square brackets ranges between $0.3$ and 1 so that $P$ is as in  
case B or perhaps an order of magnitude less. For $\delta=1$ and $n=6$, the
factor in square brackets is between 1 and 3 and so $P$ is rather larger than
in case B.   If $\delta \ll 1$ then a  
further suppression is possible.  It is worth noting that the  
reconnection probabilities found in \S7.2 were enhanced by various  
powers of $2\pi$, but that these are offset by the effects of the  
fluctuations in most cases.

For F--$(p,q)$ reconnection, $P$ contains an extra factor of $2^{k/2} /  
g_{\rm s}$, where $k = 3,6,n$ respectively in cases A,B,C.  Also the  
function $f$ is replaced with $h_{p,q}$ from eq.~(\ref{FpqProb}), which  
varies in roughly the same range as $f$ except for an extra factor of  
$q$.  Thus the F--$(p,q)$ reconnection probability is somewhat larger  
than the F--F reconnection probability, {\it if} the strings are  
coincident in the transverse directions (when the perturbative  
calculation gives $P > 1$ we assume that it is approaching saturation,  
$P \to 1$).  If the strings sit at separated minima in the transverse  
directions, the F--$(p,q)$ reconnection can be suppressed by an  
arbitrary amount, and can easily be negligible.

For the D--D reconnection probability, collecting together the  
results~(\ref{ddimp},\,\ref{ddimplarge}) and inserting numerical values  
as above, we have (roughly)
\begin{equation}
\frac{y^2}{2\pi\alpha'  \epsilon}  \sim \left\{
\begin{array}{rl}
{\rm A:}\ &\ 0.5 g_{\rm s} M\ ,\\
{\rm B:}\ &\ 15 g_{\rm s}\ ,\\
{\rm C:}\ &\  g_{\rm s} \{ O(10\ {\rm to}\ 25) + \ln{(g_{\rm  
s}/\delta)} \}\ .
\end{array}
  \right. \label{yyabc}
\end{equation}
The multiplicative contribution of each would-be fermion zero mode to  
$(1-P_{\rm pp})$ is $(1 - e^{-y^2 /2\pi\alpha'  \epsilon})$, so we see  
that these zero modes are largely lifted in all cases, and for larger  
values of $g_{\rm s}$ (but still less than 1) the production of  
fermionic open strings is negligible.  The contribution of the open  
string tachyons is $(1 + e^{\theta/\epsilon} e^{-y^2 /2\pi\alpha'   
\epsilon})$.  Defining $y^2 /2\pi\alpha'  \epsilon = g_{\rm s} K$,  
where we see that $K$ is a number of order 10 or more, the tachyon production
is suppressed for $\theta < 0.3 K  
g_{\rm s}$. Thus the suppression might be over a small range of angles  
or over all angles, depending on the precise values of $K$ and $g_{\rm s}$.

For $(p,q)$-$(p',q')$ collisions the result depends primarily on the  
values of $q$ and $q'$.  Making one or both of these larger than 1  
enhances reconnection in two ways: the production of strings is  
enhanced by the Chan-Paton degeneracy, and the fluctuations are  
decreased due to the greater tension of the $(p,q)$ string.

Note that while the naive interaction probability~(\ref{lowest}) for  
D--D collisions is close to one, we have identified two effects that can  
reduce it substantially: the need to produce $O(1/g_{\rm s})$ strings  
as discussed in \S5.5, and the fluctuation effects considered here.  We  
have seen that the former are effective only for $g_{\rm s}$ less than  
the GUT value $0.05$, while the latter become most significant as  
$g_{\rm s} \to 1$; there may be some parameter values where both play a  
role.

The reader might be concerned that in our string calculations we have  
consistently worked to lowest order in $g_{\rm s}$, but that we have  
found that higher-order effects such as the D-string fluctuations can  
have an important effect.  This is not inconsistent: the fluctuations  
are enhanced by large logarithms below the string scale, in the spirit  
of renormalization group calculations.  Also, the D--D reconnection  
probability is parametrically $e^{-O(1/g_{\rm s})}$,  but we have  
argued that the onset of this behavior is at a rather small value of  
$g_{\rm s}$.

\sect{Conclusions}

For F-strings we have found reconnection probabilities in the range  
$10^{-3}$ to 1, and for D-strings perhaps $0.1$ to 1.   With sufficient precision these can be distinguished in most cases from gauge theory strings that have  $P=1$ exactly.

If there are stable cosmic strings for more than one $(p,q)$ value, the  
reconnection probability for strings of different types plays an  
essential role.  When this is large, there is the possibility that the  
strings freeze into a three-dimensional network that is very different  
from the usual scaling solution~\cite{freeze}.  Whether this happens  
can only be determined by detailed simulations.
When the different $(p,q)$ do not reconnect, the situation would seem  
to be simpler as the different networks evolve independently.  However,  
it is not clear whether such independent networks ever form.  The  
string network likely forms in a highly tangled state, which remains as  
the different string segments roll to their respective minima.  As the  
network evolves, endpoint interactions allow the strings of different  
types to disconnect, and also allows the populations of the various  
$(p,q)$ values to change, as segments disappear and form,\footnote
{This also resolves the puzzle of what cuts off the values of $p$ and  
$q$ that are populated.  One might expect that larger values are less  
likely to form, but the usual scaling solution is an attractor so very  
large values of $(p,q)$ might begin to scale after enough time.   
However, in the tangled network there will be some equilibration.}
but it seems unlikely that this is sufficiently efficient as to allow  
them to fully disconnect.

To summarize, the following would be interesting to simulate:
\begin{enumerate}
\item
Networks of a single type of string, but with $P<1$.  For simplicity one might
initially ignore the $\theta$ and $v$ dependence and consider fixed $P$ down
to around $10^{-3}$.  The interesting question is how the various signals scale with $P$.  Ultimately one might be able to detect the
$\theta$ dependence from the spectrum of kink angles in the network, but this
is far in the future.
\item
Networks of $(p,q)$ strings, formed in an initially tangled state.  Here the interesting question is whether the scaling regime is reached.  For
simplicity one might start by taking $P$ to be exactly one for strings of the
same type, and either one or zero for strings of different types.  For recent work see refs.~\cite{pqscale}.
\end{enumerate}

Our study of string collisions has involved a wide range of interesting  
physics, both in the perturbative calculations of the reconnection  
process and in the effects of compactification.  In the best case,  
where all the $(p,q)$ strings are cosmically stable, we might hope to  
see a great deal of string physics written in the sky.

\subsection*{Acknowledgements}

We would like to thank Ed Copeland, Rob Myers, and Henry Tye for  
discussions.
MJ and NJ would like to thank the KITP for hospitality and support  
through the Graduate Fellows program.  This work was supported by  
National Science Foundation grants PHY99-07949, PHY00-98395, and  
PHY00-98631.

\end{document}